\documentclass[twocolumn]{aastex701}

\usepackage{siunitx}
\usepackage[version=4]{mhchem}

\accepted{January 6, 2026}
\submitjournal{the Astronomical Journal}

\begin{document}

\title{Two-source terrestrial planet formation with a sweeping secular resonance}
\shorttitle{Sweeping Secular Resonance during Terrestrial Planet Formation}
\shortauthors{M. Goldberg et al.}

\author[orcid=0000-0003-3868-3663]{Max Goldberg}
\affiliation{Laboratoire Lagrange, UMR7293, Universit\'{e} C\^{o}te d’Azur, CNRS, Observatoire de la C\^{o}te d’Azur, Bouldervard de l'Observatoire, 06304, Nice Cedex 4, France}
\email{maxgoldberg@oca.eu}

\author[orcid=0000-0002-4547-4301]{David Nesvorn\'{y}}
\affiliation{Solar System Science \& Exploration Division, Southwest Research Institute, 1301 Walnut St., Suite 400,  Boulder, CO 80302, USA}
\email{davidn@boulder.swri.edu}

\author{Alessandro Morbidelli}
\affiliation{Coll\`{e}ge de France, Universit\'{e} PSL, 75005 Paris, France}
\affiliation{Laboratoire Lagrange, UMR7293, Universit\'{e} C\^{o}te d’Azur, CNRS, Observatoire de la C\^{o}te d’Azur, Bouldervard de l'Observatoire, 06304, Nice Cedex 4, France}
\email{Alessandro.Morbidelli@oca.eu}

\begin{abstract}
The models that most successfully reproduce the orbital architecture of the Solar System terrestrial planets start from a narrow annulus of material that grows into embryos and then planets. However, it is not clear how this ring model can be made consistent with the chemical structure of the inner Solar System, which shows a reduced-to-oxidized gradient from Mercury to Mars and a parallel gradient in the asteroid belt. We propose that there were two primary reservoirs in the early inner Solar System: a narrow, refractory enriched ring inside of 1 au, and a less massive, extended planetesimal disk outside of 1 au with oxidation states ranging from enstatite chondrites to ordinary chondrites. We show through a suite of N-body simulations that an inwardly sweeping secular resonance, caused by aerodynamic drag and perturbations from a mean-motion resonant Jupiter and Saturn, gathers the outer planetesimal disk into a narrow ring that migrates radially, forms Mars, and contributes oxidized material to proto-Earth. Remaining unaccreted planetesimals can be implanted into the asteroid belt as the parent bodies of aubrites and non-carbonaceous iron meteorites, while the most reduced material is not implanted and thus unsampled in the meteorite collection. This model explains the oxidation and isotopic gradients within the inner Solar System within the context of a low-viscosity, magnetic wind-driven disk.
\end{abstract}


\section{Introduction}
Only in the past two decades have dynamical models finally succeeded at producing the terrestrial planets---Mercury, Venus, Earth, and Mars---with the right masses and orbits from an initial population of small objects. Although all models remain stochastic and none can match all constraints simultaneously, the most successful have some traits in common. They all rely on rapid growth of Jupiter and invoke a particular location where embryos preferentially formed. Nevertheless, several fundamental points remain contentious or unclear, including the mode of Earth's accretion \citep{Johansen2021,Morbidelli2025}, the evolution of the giant planets \citep{Izidoro2025}, and the origin of the low masses of Mercury and Mars. 

In parallel, cosmochemical studies have revealed a complex chain of events in the Solar System's first 100 Myr, including several episodes of planetesimal formation, long-distance transport of material reservoirs, and late giant impacts. These constraints have the potential to rule out dynamical hypotheses that fail to produce these distinct signatures, but drawing robust conclusions relies on proper interpretation and well-crafted models. In particular, studies of the nucleosynthetic isotope anomalies of meteorites have found a extremely diverse population of material that comprised the building blocks of the Solar System. The most fundamental result is that there are two distinct populations forming a dichotomy in almost every nucleosynthetic isotope system. The two reservoirs, referred to as non-carbonaceous (NC) and carbonaceous (CC), appear to be related to inner and outer Solar System material that was separated near the orbit of Jupiter and never mixed \citep{Budde2016,Kruijer2017}. Each reservoir also displays considerable heterogeneity among its components. In particular, the terrestrial planet mantles, enstatite chondrites, ordinary chondrites, aubrites, and some iron meteorites are all NC but vary in isotopic anomalies, formation ages, and oxidation states \citep{Krot2014,Kleine2020,Grewal2024}. Our aim in this work is to use this varied NC population to inform a dynamical model for the formation of the terrestrial planets. 

\subsection{Cosmochemical context}
Several lines of evidence suggest that the Solar System's rocky material within the orbit of Jupiter cannot arise from a single reservoir. Mercury, Earth, and Mars exhibit significant differences in both chemical and isotopic composition. In terms of oxidation state, Mercury is highly reduced \citep{Nittler2011}, Mars is highly oxidized, and Earth falls in between. Isotopically, the Earth and Martian mantles, while both clearly in the NC group, are distinct \citep{Burkhardt2021}. This diversity is mirrored in the asteroid belt. Meteorite samples show considerable variation in oxidation state between enstatite and ordinary chondrites \citep{Krot2014} and furthermore NC irons, achondrites, and chondrites span a broad isotopic range \citep{Kleine2020}. Notably, Earth is an end member in almost every isotope system and thus cannot be assembled from known meteorite populations. This challenge led \cite{Burkhardt2021} to propose that Earth formed primarily from inner disk material unsampled in the meteorite collection.

It is not obvious how to reconcile the unsampled reservoir hypothesis with the first, and to date only, dynamical models that successfully reproduce the orbital and mass architecture of the inner Solar System: those which invoke a narrow ring of material from which the planets grew \citep{Hansen2009}. Rings cannot maintain a gradient in composition because they are rapidly mixed, particularly when large embryos form within them. Earth should therefore resemble the ring's average composition and the asteroids, which in this scenario accreted from a very specific region of the ring and acquired a unique composition, should span a range around it. Earth's end member status thus requires a scenario that (a) maintains chemically distinct reservoirs from which the Earth and asteroids accrete, and (b) prevents implantation into the asteroid belt of the reservoir from which Earth dominantly grew.

Planetary mantles also provide important constraints on their accretion. \cite{Rubie2011} showed that the FeO and \ce{SiO2} fractions in Earth's mantle are not consistent with accretion from a homogeneous source of material. They argued instead that Earth initially accreted reduced material and then switched to more oxidized material after reaching about 70\% of its mass. Further work \citep{Rubie2015} implemented these equilibration models into N-body simulations of the Grand Tack, interpreting material outside 1~au as partially oxidized, and demonstrated a close match to the observed oxide abundances. \cite{Dale2025} continued this approach with an improved differentiation model and similarly found that 70--80\% of Earth's accretion must have been reduced material. Notably, \cite{Dale2025} also found that this reduced material \textit{cannot} have predominately been from enstatite chondrites, but was instead enriched in refractory elements beyond any known NC chondrites. This is additional support of the lost reservoir hypothesis independent of the isotopic evidence from which it was originally proposed. 

Finally, the Hf-W chronometer gives information on the core formation timescale of Earth and Mars. Mars formed quickly, either entirely by 2--4~Myr if smooth accretion is assumed \citep{Dauphas2011}, or up to 15 Myr in a model with large projectiles \citep{Marchi2020}. In contrast, the final stages of Earth's accretion (consisting primarily of giant impacts) occurred in 50--100~Myr \citep{Rudge2010}. 

\subsection{Two source models}
A two source model promises to fulfill our requirements. We propose that the inner ring was chemically very reduced and refractory enriched and corresponded isotopically to the unsampled NC reservoir proposed in \cite{Burkhardt2021}. It would provide most of the material to form Mercury, most of Venus and Earth, and be entirely accreted onto the terrestrial planets rather than implanted in the asteroid belt. The outer reservoir, more oxidized, would then be the source of the NC iron meteorite parent bodies (see Section~\ref{sec:swept}) and most of Mars, contribute to the growth of Earth, and eventually form ordinary chondrites. 

\cite{Nesvorny2025} performed numerous N-body integrations to test this two source scenario. Several of their models showed excellent ability to replicate the orbital architecture of the Solar System as evaluated by the presence of a small Mercury and Mars in their proper location and a large and closely-spaced Venus and Earth. Their most successful models are characterized by several distinct features. First, a very massive gas disk with a broad pressure bump at 1~au provides convergent migration that keep Venus and Earth close to each other. Second, the inner ring is narrow and located at $\sim 0.5$ au, so that large embryos migrate outwards to become Venus and Earth but the last-forming embryo is stranded close in and becomes Mercury. Third, the outer source is also a confined ring of mass 30--50\% of the inner ring and located near 1.5--2 au. 

A major question in this scenario is the origin of the outer planetesimals. Although the inner ring is likely associated with the silicate sublimation line and/or the MRI dead zone transition \citep{Marschall2023}, no prominent condensation line or obvious disk structure is present near 1.7 au. A narrow, compositionally-uniform ring is also disfavored by the considerable isotopic diversity of NC achondrites which furthermore do not even show an age--isotope anomaly correlation \citep{Kleine2020}. For this reason, we think the most likely scenario is that the outer reservoir was a wide and long-lived dust disk that underwent sporadic planetesimal formation spaced in heliocentric radius and time. \cite{Lenz2019} showed that vortices that appear and disappear stochastically in many regions in the disk can concentrate dust until the dust collapses gravitationally and forms a planetesimal. A physical mechanism in the disk must then be invoked to concentrate this dispersed planetesimal population into a ring at $\sim$1.5--1.7 au in order to form Mars and deliver some material to the accreting Earth. We turn to these dynamics now.

\subsection{Sweeping secular resonances}
Embryos and planetesimals in the protosolar disk have a complex set of dynamics driven by interactions with the gaseous disk, interactions with distant giant planets, and self-excitation. Importantly for this study, torques from the massive gas disk and the giant planets drive apsidal precession of planetesimals in the disk. Commensurability between these planetesimals' apsidal precession rate and the precession of a distant planet, known as a secular resonance, leads to an eccentricity pumping of the planetesimal and its rapid inward migration via aerodynamic drag. As the rate of precession depends on both the semi-major axis of the planetesimal and the density of the gas disk, the location of the secular resonance varies as the disk dissipates and can in principle sweep over a large region of the inner Solar System.

The theory of sweeping secular resonances has been developed extensively in the literature. They were first described by \cite{Heppenheimer1980} and \cite{Ward1981a} in the context of exciting the eccentricities of asteroids. Later \cite{Lecar1997} realized that asteroids with eccentricities pumped by a secular resonance would experience enhanced aerodynamic drag and rapidly spiral into the inner solar system, clearing the asteroid belt and preventing a planet from forming there. These calculations were expanded extensively by \cite{Nagasawa2000,Nagasawa2005} and \cite{Thommes2008}, who showed that certain gas disk models can both leave behind an appropriately excited asteroid belt and also trigger constructive growth in the terrestrial planet region. More recently, \cite{Bromley2017} attempted to link this process to the rapid formation of Mars and concluded that the dynamical excitation led to fragmentational collisions that suppressed planet formation beyond 2 au.

Nevertheless, the hypothesis that sweeping secular resonances shaped the inner Solar System has also faced criticism. The most questionable assumption is that most authors have taken the giant planets' orbits to be the same as they are today. To the contrary, several lines of evidence suggest that the disk-phase orbits of the giant planets were significantly different. First, Jupiter and Saturn should have undergone Type II migration in the disk and likely been captured into a mean motion resonance with each other \citep{Masset2001,Morbidelli2007,Walsh2011}, in stark contrast to their current non-resonant configuration. Second, many features of the Solar System are well-explained by a significant evolution of the orbital elements of the giant planets during a dynamical instability \citep{Tsiganis2005}. \cite{OBrien2007} reconsidered secular resonance sweeping in this context by assuming that Jupiter and Saturn were on near-circular, closely-spaced orbits and found that the secular resonances are both too weak to excite the asteroid belt and do not extend far enough inwards to affect the terrestrial planets.

Our approach is different than previous work in two major ways. First, we consider the effect of the secular resonance in producing the chemical composition of the terrestrial planets and the asteroid belt in the context of the two-source model described above. We track the initial location of planetesimals and the resulting composition of the final embryos, for which constraints exist in the Solar System. Second, we take as assumptions results derived from modern simulations of low viscosity disks in the context of the Solar System. Specifically, our gas disk profile is peaked at 1 au, inspired by simulations of disks where angular momentum transport is dominated by magnetic disk winds \citep{Suzuki2016,Kunitomo2020}. We further assume that Jupiter and Saturn have been trapped in 2:1 mean motion resonance due to slow Type II migration in the disk \citep{Griveaud2023,Griveaud2024}. These choices and their consequences will be described in detail below.

\section{Sweeping secular resonances}
\label{sec:ssr}
Suppose a planetesimal $p$ is orbiting interior to Jupiter with semi-major axis $a_p$ and eccentricity $e_p$. It experiences a secular potential from Jupiter and Saturn as well as the disk gravity. The total perihelion precession rate of this planetesimal, $g_p = \dot{\varpi}_p$, can be decomposed into three main parts,
\begin{equation}
    g_p = g_{p,\textrm{disk}} + g_{p,\textrm{J}} + g_{p,\textrm{S}}
\end{equation}
where $g_{p,\textrm{J}}$ and $g_{p,\textrm{S}}$ are the precession driven by Jupiter and Saturn and $g_{p,\textrm{disk}}$ is due to the disk. The disk-driven precession $g_{p,\textrm{disk}}$ is given by
\begin{equation}
    g_{p,\textrm{disk}} = \frac{2}{n_pa_p^2}T(a_p)
\end{equation}
where $n_p$ is the mean motion of the planetesimal, and
\begin{multline}
    T(a) = \frac{1}{4} \iiint\biggl[\frac{-3 a + 2 r \cos \phi}{(a^2+r^2+z^2-2ar\cos\phi)^{3/2}} \\+ \frac{3z^2}{(a^2+r^2+z^2-2ar\cos\phi)^{5/2}} \biggr] G\rho(r,\phi,z) dV,
\end{multline}
which we solve by numerical integration \citep{Nagasawa2000}. Importantly, $g_{p,\textrm{disk}}$ depends on the specifics of the disk profile, but in general is proportional to the local gas density $\rho$ and negative for realistic disks. 

The narrow confinement of mass in the inner Solar System has motivated the hypothesis that the Solar disk had a pressure peak near 1 au \citep{Ogihara2018}. Indeed, \cite{Woo2023} showed that disks with a canonical power law slope, e.g., the Minimum Mass Solar Nebula (MMSN), rapidly lose planets to the inner disk by Type I migration. They showed instead that a peaked disk provides convergent migration and reproduces the mass profile of the Solar System \citep{Woo2024}. \cite{Nesvorny2025} performed extensive testing of this family of models, focusing especially on their ability to produce two closely-spaced massive planets and a small Mercury and Mars. We adopt their most successful surface density profile,
\begin{equation}
    \Sigma_\textrm{g}(r,t) = \Sigma_0 \left(\frac{r}{r_0}\right)^{-\ln(r/r_0)} \exp(-t/\tau_0)
    \label{eq:gas}
\end{equation}
where $r_0=\qty{1}{au}$, $\Sigma_0=\qty{3000}{g.cm^{-2}}$, and the disk dissipation timescale $\tau_0$ is a parameter of each simulation. The volumetric gas density is given by
\begin{equation}
    \rho(r,z,t) = \frac{1}{\sqrt{2\pi}} \frac{\Sigma}{H} \exp(-z^2/(2H^2))
\end{equation}
and the disk is flared, with aspect ratio
\begin{equation}
    H/r = 0.03358 (r/r_0)^{1/4}
    \label{eq:flaring}
\end{equation}
and the disk is assumed to extend from $r=\qty{0.1}{au}$ to $\qty{4.0}{au}$, where it is truncated by Jupiter. 

This gas disk profile has several important features. Its broad shape closely resembles that predicted in magnetic disk wind models \citep{Suzuki2016} with a peak at 1 au. Steep falloff on both sides of the peak cause strong convergent migration towards 1 au. Notably, outward migration of material growing at 0.5 au can strand small embryos which later become Mercury \citep{Nesvorny2025}. We adopt a more massive Solar Nebula than the canonical $\Sigma_0 = \qty{1700}{g.cm^{-2}}$ \citep{Hayashi1981} because $\Sigma_0=\qty{3000}{g.cm^{-2}}$ was shown by \cite{Nesvorny2025} to work better for terrestrial planet formation by increasing the strength of migration out of the inner ring and reducing the final Venus-Earth separation distance. The disk is also assumed to dissipate uniformly in time. While more complex depletion models can have slightly different behavior \citep{Nagasawa2000}, in the absence of a complete model of disk evolution incorporating magnetic winds, photoevaporation, and the effect of multiple gas giants, we consider only the simplest case. 

Now we turn to the giants. The precession of a planetesimal caused by a planet with semi-major axis $a_\textrm{planet}$ and mass $m_\textrm{planet}$ is, to first order in eccentricity, 
\begin{equation}
    g_{p,\textrm{planet}} = \frac{1}{4} \alpha^2 b_{3/2}^{(1)} \frac{m_\textrm{planet}}{M_*} n_p
    \label{eq:gp}
\end{equation}
where $n_p$ is the mean motion of the planetesimal, $M_*$ is the mass of the Sun, and $b_{3/2}^{(1)}$ is the Laplace coefficient that is a function of $\alpha=a_p/a_\textrm{planet}$. Note that Equation \ref{eq:gp} depends only on the mass and semi-major axis of the giant planet, and is independent of its eccentricity as well as the eccentricity of the planetesimal. Furthermore, secular precession is always forward, i.e., $g_{p,\textrm{planet}} > 0$. 

Secular resonances occur when the precession rate of a planetesimal equals one of the frequencies governing the precession rate of Jupiter or Saturn. In the non-resonant case, there are two such frequencies, $g_5$ and $g_6$, which arise as eigenfrequencies in the Laplace-Lagrange solution. The resonance condition is then $g_p = g_5$ or $g_p = g_6$. In secular resonance, the time-averaged secular forcing from the giant planets grows dramatically and the planetesimal acquires a large eccentricity. Because $g_{p,\textrm{disk}}$ depends on the gas density, which is expected to evolve over the lifetime of the disk (and eventually reaches zero), the locations of the secular resonances change during the evolution of, and especially the dissipation of, the disk. 

Many previous studies of secular resonance sweeping in the Solar System assumed that Jupiter and Saturn were on their contemporary orbits during the protoplanetary disk phase \citep{Lecar1997,Nagasawa2000,Nagasawa2005,Thommes2008}. Under this assumption, the precession of the giant planets is always forward and with the current rates $g_5$ and $g_6$. Consequently, the secular resonances approach their current locations as the gas fully dissipates \citep{Heppenheimer1980}. However, as discussed in the Introduction, there is considerable evidence that the orbits of the giant planets were initially set by planet-disk interactions and changed early in the Solar System's evolution. Some works have considered alternate starting locations for Saturn but nevertheless use non-resonant orbits \citep{OBrien2007} or Laplace-Lagrange secular theory \citep{Bromley2017}, which fails in proximity to mean-motion resonance. Thus, it is critical to reexamine the theory of sweeping secular resonances under more probable versions of the disk-phase orbits of the giant planets.

For these reasons, we will focus on the secular behavior of Jupiter and Saturn on mean-motion resonant orbits that would realistically result from planet migration. Early hydrodynamical work found that Jupiter and Saturn frequently capture into 3:2 mean motion resonance during Type II migration \citep{Masset2001,Morbidelli2007}, and indeed this configuration can evolve into the current giant planet orbital architecture \citep{Walsh2011,Nesvorny2012a}. However, more recent hydrodynamical studies using smaller disc viscosities, believed to be more realistic for the MRI-inactive region of the disk, found that Jupiter and Saturn consistently capture into the 2:1 resonance and never 3:2 because they migrate more slowly \citep{Pierens2014,Griveaud2023,Griveaud2024}. \cite{Griveaud2024} also found that some four- and five-giant planet resonant chains formed in hydrodynamical simulations could undergo dynamical instability and evolve into the current Solar System architecture. Accordingly, we take their C4 and C5 configurations of Jupiter and Saturn as our two giant planet configurations, labeled REJSC4 and REJSC5, respectively.

The mechanism of secular resonance sweeping is more complicated if Jupiter and Saturn are in resonance for two reasons. First, planets in mean motion resonance precess backwards, rather than forwards as in the non-resonant case. Therefore, for a planetesimal to be caught in secular resonance, it must initially precess backwards, $g_{p}<0$, as well. As $g_{p,\textrm{J}}$ and $g_{p,\textrm{S}}$ remain positive (the planetesimal still feels Jupiter and Saturn secularly, as averaged rings of mass), $g_{p,\textrm{disk}}$ must be larger in absolute value so that the total $g_p$ is negative. In the non-resonant Jupiter-Saturn case, the secular resonances exist even without a disk and simply move closer to Jupiter with more gas. In the resonant case, on the other hand, the secular resonances do not exist at all without gas. A minimum gas density is therefore required for secular resonances to appear and their locations are more sensitive to the gas disk profile.

The second complication is that the precession rate of a mean-motion resonant Jupiter and Saturn is dependent on their eccentricity, scaling approximately as $1/e$, in contrast to the non-resonant case where it is nearly independent of $e$. So, the precession cannot be uniquely determined just from the 2:1 resonant orbit and needs to be calculated knowing all of the orbital elements. We compute it by numerically integrating Jupiter and Saturn in their starting configuration. As in the non-resonant case, multiple frequencies are present in the evolution of the eccentricity vectors of Jupiter and Saturn, but in practice the precession of Jupiter is dominated by a single, negative frequency in both REJSC4 and REJSC5 that we report as $g_\textrm{J}$.

These two cases nevertheless differ in several ways despite having the same semi-major axes of Jupiter and Saturn. In REJSC4, $e_\textrm{J}\approx 0.08$ and $e_\textrm{S}\approx 0.14$, the relative argument of pericenter $\varpi_\textrm{J}-\varpi_\textrm{S}$ is librating with small amplitude, and $g_\textrm{J} \approx 2\pi/(\qty{-6494}{yr})$. In REJSC5, $e_\textrm{J}\approx 0.11$, $e_\textrm{S}\approx 0.06$, $\varpi_\textrm{J}-\varpi_\textrm{S}$ is circulating, and $g_\textrm{J}\approx 2\pi/(\qty{-9805}{yr})$. The circulation of the relative pericenters causes the eccentricity vector of Saturn to have two modes comparable in amplitude, of frequency $2\pi/(\qty{-26318}{yr})$ and $2\pi/(\qty{-6025}{yr})$. In practice, the secular resonance sweeping effect is determined almost solely by $g_\textrm{J}$ and $e_\textrm{J}$, the former setting its location and the latter its strength. 

\begin{figure}
    \centering
    \includegraphics[width=\linewidth]{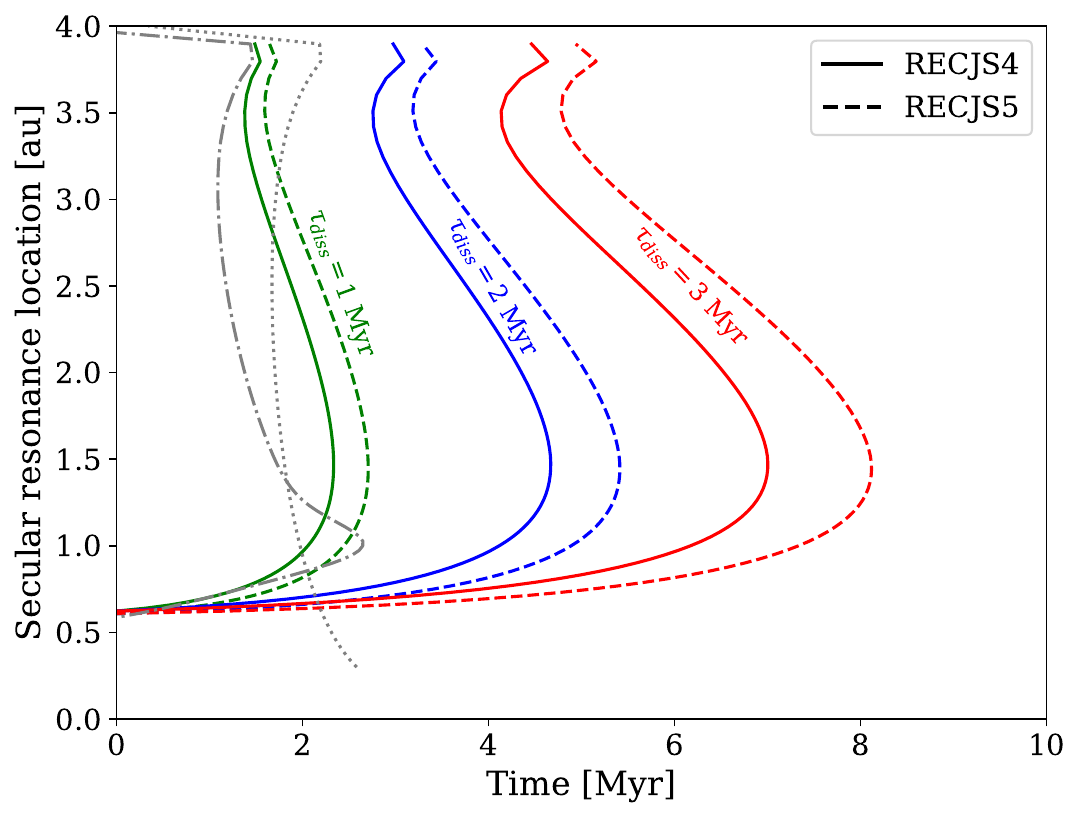}
    \caption{The locations of the secular resonances as a function of time during disk dissipation with the gas disk profile used in this work. Colors represent the three disk dissipation timescales $\tau_\textrm{diss}=\{1,2,3\}$ Myr and solid and dashed lines are the two Jupiter-Saturn configurations. The gray dotted and dash-dotted lines are a $1/r$ and broken-power law \citep{Woo2023} gas profile, respectively, but otherwise assume the same parameters as the solid green line.}
    \label{fig:secreslocs}
\end{figure}

Figure \ref{fig:secreslocs} shows the location of the $g_\textrm{J}$ secular resonances as a function of time for $\tau_\textrm{diss}=\{1,2,3\}$ Myr and REJSC4 and REJS5. We also plot, for comparison, the secular resonance locations for a $\Sigma_g\propto 1/r$ disk model and for the ``shallow inner'' profile used in \cite{Woo2023,Woo2024} (dotted and dashed-dotted line, respectively), both assuming $\tau_\textrm{diss}=\qty{1}{Myr}$, REJSC4, and $\Sigma_0=\qty{3000}{g.cm^{-2}}$. An important feature seen in Figure \ref{fig:secreslocs} is that for all of these parameters, the secular resonance exterior to $\sim \qty{1.5}{au}$ evolves inwards over time. This property is not guaranteed for negatively precessing Jupiter-Saturn, and in fact is absent for disk profiles that do not fall off steeply enough outside of 1 au. However, this is the critical piece that allows planetesimals to ``surf'' on a secular resonance: as their eccentricities are excited by the resonance and they migrate inwards by aerodynamic drag, the secular resonance catches up to them to repeat the process, enabling long-distance migration (provided their aerodynamic drag is faster than the evolution of the secular resonance \citealt{Nagasawa2005}). In contrast, an outwardly-evolving secular resonance will only excite the planetesimals a single time, whereupon they aerodynamically damp back to circular orbits and only briefly migrate inwards. Hence, the branch between 0.5 and 1.5 au, due to gas depletion in the inner disk and visible near the bottom of Figure~\ref{fig:secreslocs}, has almost no impact on the dynamics.

Finally, we note that if Jupiter and Saturn were caught in the 3:2 resonance, as suggested by \cite{Walsh2011} and others, the negative precession of Jupiter and Saturn would be much more rapid, on the order of $g_\textrm{J,S} \sim 2\pi/(\qty{-150}{yr})$. In this case, reasonable disk models will never drive fast enough negative precession on planetesimals to cause a secular resonance. The results of this work are thus dependent on Jupiter and Saturn being caught in a 2:1 or similarly weak mean motion resonance.

\section{Methods}
To model terrestrial planet accretion with the effect of sweeping secular resonances, we performed a suite of N-body simulations using the GPU-accelerated hybrid-symplectic code GENGA \citep{Grimm2014,Grimm2022}. GENGA parallelizes the force calculations and Kepler steps across the large number of cores in a GPU, allowing for efficient integrations with a large number of fully interacting particles. 

Following the most successful two source model from \cite{Nesvorny2025}, we initialized our simulations with two populations of planetesimals: an inner planetesimal ``ring'' and an outer planetesimal ``disk.'' Each planetesimal had mass $\qty{3.16e{-9}}{M_\odot}=\qty{6.29e24}{g}$. The inner population had a total mass of $\qty{1.6}{M_\oplus}$, corresponding to 1520 particles. Planetesimals in the inner group had their semi-major axis drawn from a Gaussian distribution with mean $\qty{0.5}{au}$ and standard deviation $\qty{0.05}{au}$ to form a narrow ring. The outer group had a total mass of $M_\textrm{out}=\qty{0.5}{M_\oplus}$ or $\qty{0.8}{M_\oplus}$ depending on the simulation, corresponding to 475 or 760 particles, respectively. Outer planetesimal semi-major axes were drawn uniformly from 1 au to 3 au. Particles in both populations had their eccentricities and inclinations drawn from Rayleigh distributions with scales of $10^{-3}$ and $0.5 \times 10^{-3}$ rad, respectively, and their longitudes of ascending node, arguments of pericenter, and initial mean anomalies were drawn uniformly from 0 to $2\pi$. Jupiter and Saturn were placed in 2:1 resonance with orbital elements found by \cite{Griveaud2024} after they smoothly removed the gas causing migration. We rescaled their orbits so that Jupiter's initial semi-major axis is 5.2 au. As they ran 2D hydrodynamic simulations, we assumed zero inclinations for Jupiter and Saturn. This has the effect of neglecting inclination-type secular resonances \citep{Ward1981a}.

Following \cite{Woo2023,Woo2024}, we modified GENGA in several ways. Our integrations included the effect of the gaseous disk with parameters given by Eqs.~\ref{eq:gas}--\ref{eq:flaring}. All particles except the Sun and the giant planets felt forces of Type I migration according to the formulae of \cite{Paardekooper2011} and as implemented in \cite{Ogihara2018}, including Lindblad and corotation torques and the effect of saturation. We also include tidal damping of eccentricity and inclination according to \cite{Cresswell2008}. Aerodynamic drag and the gravitational potential of the disk are applied to planetesimals and embryos, as previously implemented in GENGA using the formalism of \cite{Morishima2010}. We neglect the effect of the disk potential on the giant planets to maintain consistency with the post-gas resonant Jupiter-Saturn configurations that we use from \citep{Griveaud2024}. In reality, owing to the large mutual gap surrounding the giants, they would feel a small forward precession about an order of magnitude slower than that felt by planetesimals at 3 au, only slightly shifting the position of the secular resonance.

Finally, following \cite{Woo2023}, we included a simple superparticle algorithm in which the original particles are intended to each represent a cloud of 4000 planetesimals each of diameter 100 km. Pairwise gravitational interactions between superparticles are reduced in strength by a factor of $1/\sqrt{4000}$ to account for the mass dependence of viscous stirring \citep{Morbidelli2009}. Aerodynamic drag, collisions, and Type I migration are calculated for superparticles as if they had a diameter of $\qty{100}{km}$ and corresponding mass assuming a density of $\qty{3}{g.cm^{-3}}$. Superparticles that collide are immediately upgraded to normal particles whose gravitational interactions, Type I migration, aerodynamic drag, and collisions are calculated as usual assuming their true mass and a radius corresponding to density $\qty{3}{g.cm^{-3}}$. 

We studied several cases by varying key parameters. The disk dissipation timescale $\tau_\textrm{diss}$ was chosen from 1, 2, or 3 Myr, the mass of the outer annulus was $\qty{0.5}{M_\oplus}$ or $\qty{0.8}{M_\oplus}$, and we tested both Jupiter-Saturn configurations REJSC4 and REJSC5. Because the success of even the best terrestrial planet accretion models remains highly stochastic \citep{Nesvorny2025}, each parameter set was simulated 5 times, to make a total of 60 simulation runs. All integrations used a constant timestep of $\qty{4.87}{d} = \textrm{yr}/75$ and were run until $t=\qty{15}{Myr}$. 

\section{Results}
\begin{figure}
    \centering
    \includegraphics[width=\linewidth]{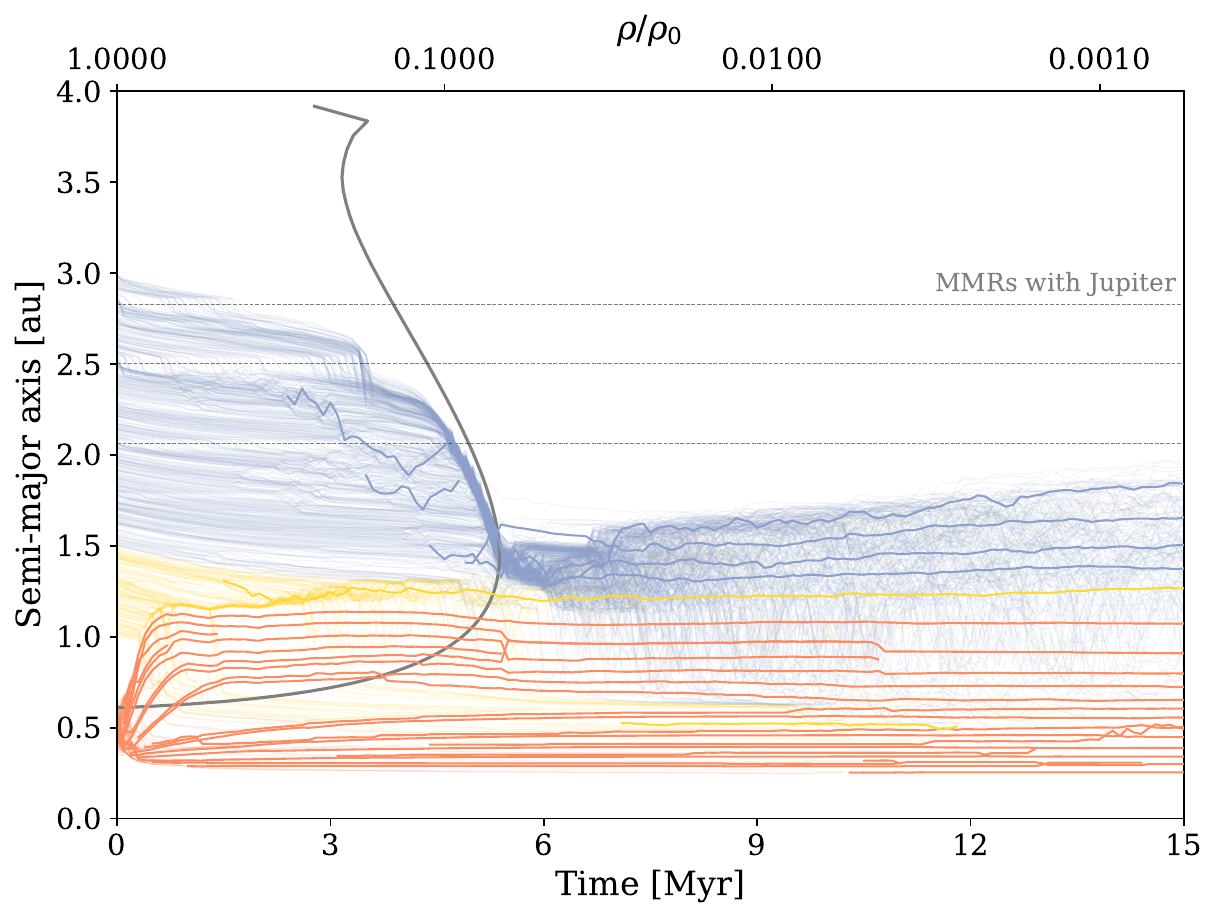}
    \caption{Typical evolution of one of our simulations, in this case using $M_\textrm{out}=\qty{0.8}{M_\oplus}$, $\tau_\textrm{diss}=\qty{2}{Myr}$ and REJSC5. Faint lines represent planetesimals and thick lines embryos (i.e., products of mergers). Lines are colored according to the planetesimal starting location. We propose that the orange and yellow particles have a reduced composition, while blue particles are oxidized, although the boundary between reduced and oxidized may not be sharp, nor does it necessarily coincide with the boundary between the swept and unswept planetesimal disks. The curved gray line marks the location of the secular resonance with Jupiter that sweeps over the inner Solar System during depletion of the gas, triggering inward migration of planetesimals and embryos. Dashed gray lines are the 3:1, 5:2, and 4:1 mean motion resonances with Jupiter.}
    \label{fig:tracks}
\end{figure}

\begin{figure*}
    \centering
    \includegraphics[width=\linewidth]{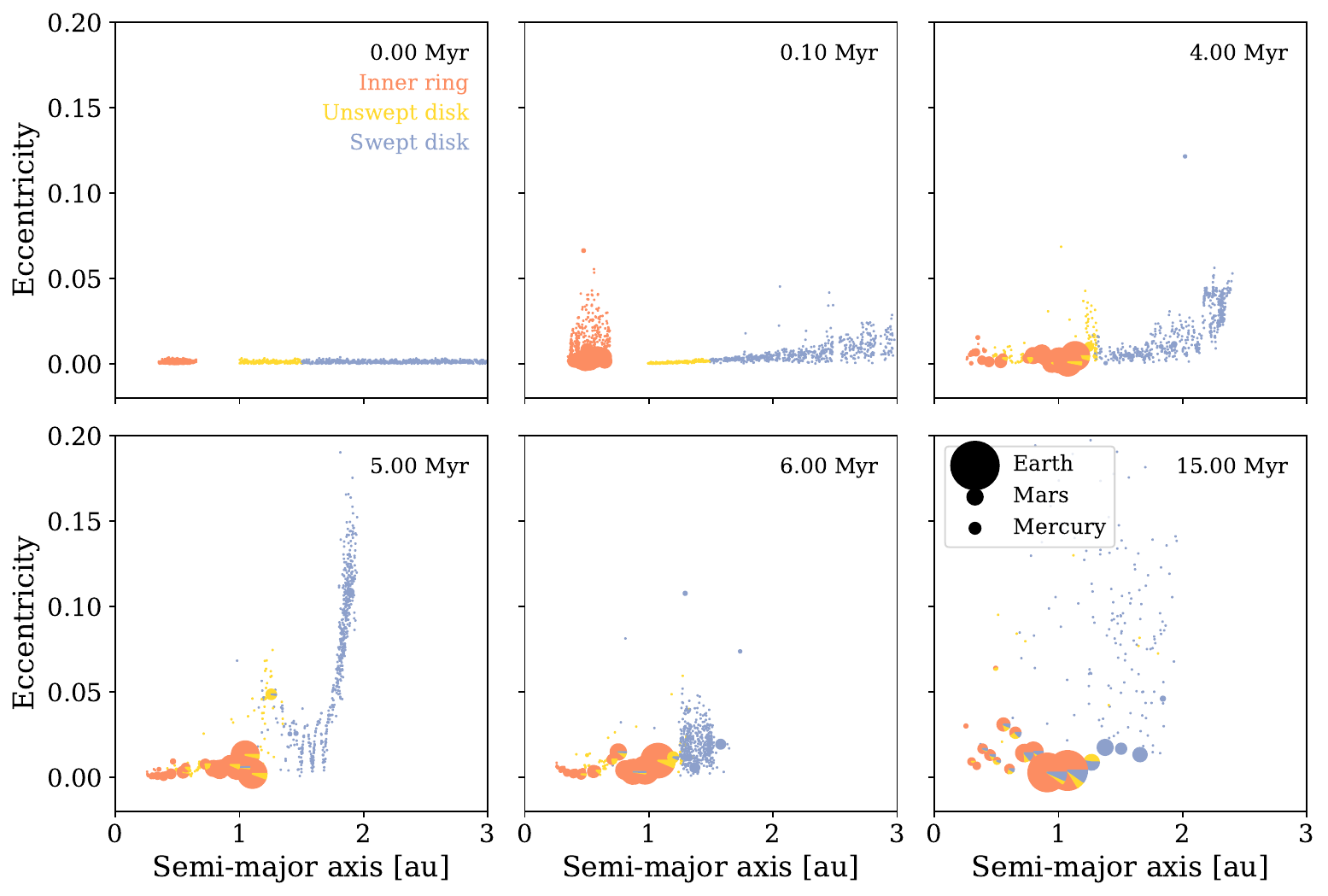}
    \caption{The same simulation as in Figure~\ref{fig:tracks} but now in $a$-$e$ space. Each embryo is a pie chart representing its relative fraction of inner and outer disk material and sized according to its mass. The inner ring rapidly grows embryos that move outwards to 1 au by Type I migration. Later, the sweeping secular resonance with aerodynamic drag collects the spread-out population of outer disk planetesimals into an eccentric ring and deposits them near 1.5 au where they rapidly grow into Mars-sized embryos. }
    \label{fig:avse}
\end{figure*}

Figures~\ref{fig:tracks} and \ref{fig:avse} show the typical outcome of one of our simulations. In the inner ring, our simulations proceed similarly to those of \cite{Nesvorny2025}. Several large embryos grow quickly and migrate outwards to the migration trap at 1 au. The slowest growing embryos do not become massive enough to feel a significant Type I torque before the disk dissipates and they remain stranded at 0.5 au through the end of the simulation. This growth scenario was found to be particularly effective at forming good Mercury, Earth, and Venus \citep{Clement2021,Nesvorny2025}.

The behavior in the outer region is very different from previous simulations. Initially, there is almost no growth because the surface density is low and aerodynamic drag keeps planetesimals on circular, non-crossing orbits. The only evolution visible in Figure~\ref{fig:tracks} is that planetesimals slowly drift inwards due to drag and Kirkwood gaps are cleared. Upon the passage of the secular resonance through this extended disk (occurring at $t\approx 2\tau_\textrm{diss}$ for our choice of disk parameters), all planetesimals receive a large boost in eccentricity and begin to rapidly migrate inwards. Planetesimals that migrate faster than the inward evolution of the secular resonance lose their eccentricity and stop migrating. This creates a ``bunching'' effect where almost all of the particles that started outside of 1.5 au converge to the same orbit, i.e., a narrow ring that is slightly eccentric ($e\sim 0.1$) and periapse-aligned \citep{Best2024}. In essence, the sweeping secular resonance provides a natural way to generate---from almost any distribution of planetesimals beyond 1 au---the second ring hypothesized in \cite{Nesvorny2025}, which was shown to be effective in forming Mars but not well-motivated from a planetesimal formation standpoint.

The final position of this ring primarily depends on $\tau_\textrm{diss}$. Slower disk dissipation slows the inward evolution of the secular resonance, making it easier for planetesimals to migrate at the same rate. In general, the semi-major axis of the ring at the time that it detaches from the secular resonance is between 1.3 and 1.7 au. Dramatically increased surface densities during the sweeping cause rapid growth into embryos and the ring viscously spreads quickly after circularizing. Importantly, for rings that land at $\lesssim 1.5$ au, a significant amount of oxidized material can accrete on the proto-Earth and Venus, which we discuss in Section~\ref{sec:chem}.

Planetesimals that started between 1 and 1.5 au have a distinct trajectory to those that started outside of 1.5 au. First, many of them are scattered inwards of 1 au when the largest embryos from the inner ring migrate outwards. The remainder are left near their original location but are not significantly affected by the sweeping secular resonance because it is evolving outwards with time interior to 1.5 au (see Section~\ref{sec:ssr}). Because of the very different outcomes of disk planetesimals that originate exterior and interior to 1.5 au, we track these populations separately and refer to them as the ``swept disk'' and ``unswept disk,'' respectively.

As we are addressing the effects of the dynamics of terrestrial planet formation on the eventual chemical composition of the planets and remaining small bodies, we must discuss both the orbital and chemical architectures established by our models. We begin with widely-used criteria to evaluate the success of terrestrial planet models in producing a set of planets that resembles the inner Solar System in mass and orbits. Because our simulations end at 15 Myr and before the giant planet instability, we do not expect their final states to precisely match the terrestrial planets, especially Earth, which continued accreting for $\sim 100$ Myr. Instead, we will compare our simulations to the most successful models of \cite{Nesvorny2025} sampled at comparable times, i.e. predating the giant planet instability.

\subsection{Mass architecture}
\begin{figure}
    \centering
    \includegraphics[width=\linewidth]{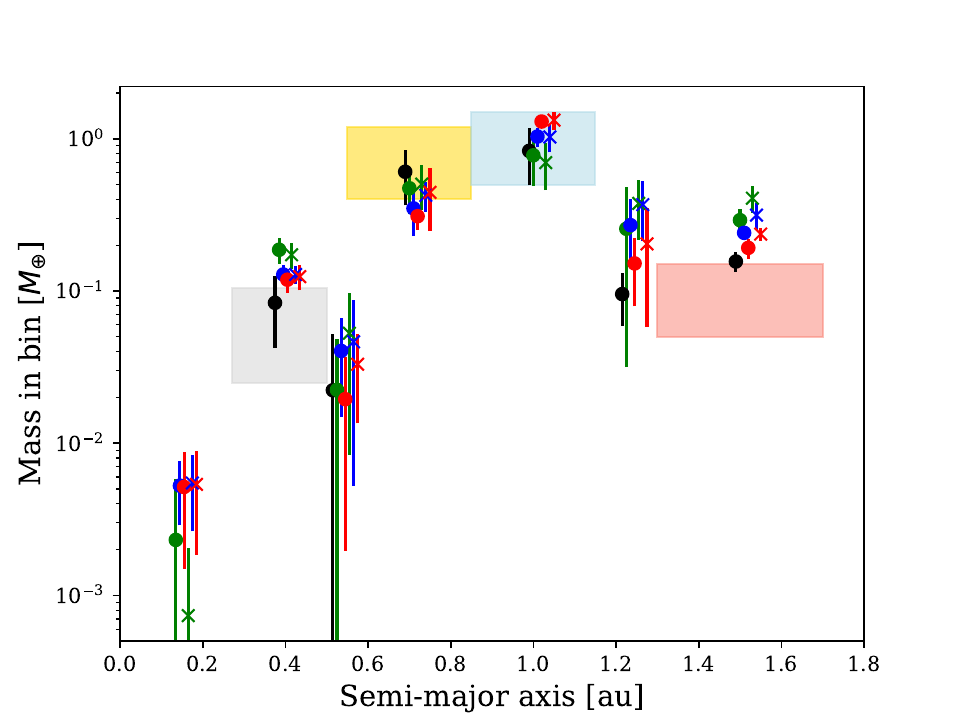}
    \caption{Mass in each planet bin at the end of each simulation. Black points are from \cite{Nesvorny2025} at 5 Myr. Colored points are this work. Circles and crosses are $M_\textrm{out}=\qty{0.5}{M_\oplus}$ and $\qty{0.8}{M_\oplus}$ respectively. Green, blue, and red are $\tau_\textrm{diss}=$ 1, 2, and 3 Myr, respectively.}
    \label{fig:massdist}
\end{figure}

Of primary concern is the spatial distribution of mass interior to Jupiter. Many models struggle to produce the small masses of Mercury and Mars and the close spacing of Earth and Venus. To quantify these, we first compute the mass contained in the bins $[0.27, 0.5]$, $[0.55, 0.85]$, $[0.85, 1.1]$, and $[1.3, 1.7]$ au, as these are expected to accrete efficiently onto Mercury, Venus, Earth, and Mars, respectively \citep{Brasser2016}. The binned mass distribution is shown in Figure~\ref{fig:massdist} compared to the simulations of \cite{Nesvorny2025} at 5 Myr.

Furthermore, we compute the radial mass concentration, defined as
\begin{equation}
    S_c = \max\left[\frac{\sum_j m_j}{\sum_j m_j \left[\log_{10}(a/a_j)\right]^2}\right]
\end{equation}
where $m_j$ and $a_j$ are the mass and semi-major axis of particle $j$ and the maximum is taken over $a$ \citep{Chambers2001}. 
Higher $S_c$ indicates a narrower mass distribution.

Values of $S_c$ in our simulations are given in Table~\ref{tab:stats} averaged over the two Jupiter-Saturn configurations and five runs with different initial conditions. 
For comparison, \cite{Nesvorny2025} simulations have $S_c = 69.0\pm 18.8$ at 5 Myr and the real Solar System has $S_c = 89.9$.
The most significant trend is that $S_c$ is larger for higher $\tau_\textrm{diss}$ because the outer ring is deposited deeper in the terrestrial planet region. Higher $M_\textrm{out}$ also leads to lower $S_c$ because more mass is present in the outer ring.

\begin{deluxetable*}{ccccccc}
\caption{Radial mass concentration ($S_c$) and fraction $p$ of initial particles from the three populations that were not accreted by the end of the integration\label{tab:stats} among our simulation sets}
\scriptsize
 \tablehead{
 \colhead{$\tau_\textrm{diss}$ (Myr)} & \colhead{$M_\textrm{out}~(M_\oplus)$} & \colhead{$S_c$ at 5 Myr} & \colhead{$S_c$ at 15 Myr} & \colhead{$p_\textrm{inner}$} & \colhead{$p_\textrm{unswept}$} & \colhead{$p_\textrm{swept}$}
 }
\startdata
\tableline
1 & 0.5 & $46.0 \pm 4.3$ & $39.2 \pm 3.3$ & $(1.5 \pm 1.2) \times 10^{-3}$ & $0.37 \pm 0.12$ & $0.70 \pm 0.10$ \\
1 & 0.8 & $43.2 \pm 2.6$ & $35.1 \pm 1.4$ & $(9.2 \pm 9.4) \times 10^{-4}$ & $0.25 \pm 0.08$ & $0.57 \pm 0.10$ \\
2 & 0.5 & $57.7 \pm 4.8$ & $48.9 \pm 5.1$ & $(1.6 \pm 1.6) \times 10^{-3}$ & $0.14 \pm 0.08$ & $0.37 \pm 0.08$ \\
2 & 0.8 & $56.8 \pm 2.5$ & $46.2 \pm 2.8$ & $(1.1 \pm 1.0) \times 10^{-3}$ & $0.042 \pm 0.023$ & $0.24 \pm 0.03$ \\
3 & 0.5 & $64.0 \pm 5.7$ & $62.3 \pm 3.9$ & $(2.8 \pm 2.0) \times 10^{-3}$ & $0.12 \pm 0.06$ & $0.29 \pm 0.08$ \\
3 & 0.8 & $60.5 \pm 4.0$ & $60.3 \pm 7.6$ & $(1.5 \pm 0.9) \times 10^{-3}$ & $0.03 \pm 0.02$ & $0.20 \pm 0.05$ \\
\enddata
\end{deluxetable*}

In general, our typical $S_c$ values between 40 and 60 are well below the real Solar System value. As can be seen in Figure~\ref{fig:massdist}, this is primarily due to an excess of mass near 1.5 au and the formation of several, rather than one, Mars-sized embryos. We anticipate that during the giant planet instability some of this material will be accreted onto Earth and some will be ejected from the terrestrial region, landing in the asteroid belt or continuing past Jupiter. Although there is a considerable dependence on the specifics of the giant planet orbits during the instability, this clearing likely happens preferentially beyond 1 au because of rapid forward precession near (but outside) the 2:1 Jupiter-Saturn resonance that moves the $g_5$ secular resonance into the 1--3 au region \citep{Brasser2009}.

\subsection{Chemical contribution to terrestrial planets}
\label{sec:chem}
\begin{figure*}
    \centering
    \includegraphics[width=0.7\linewidth]{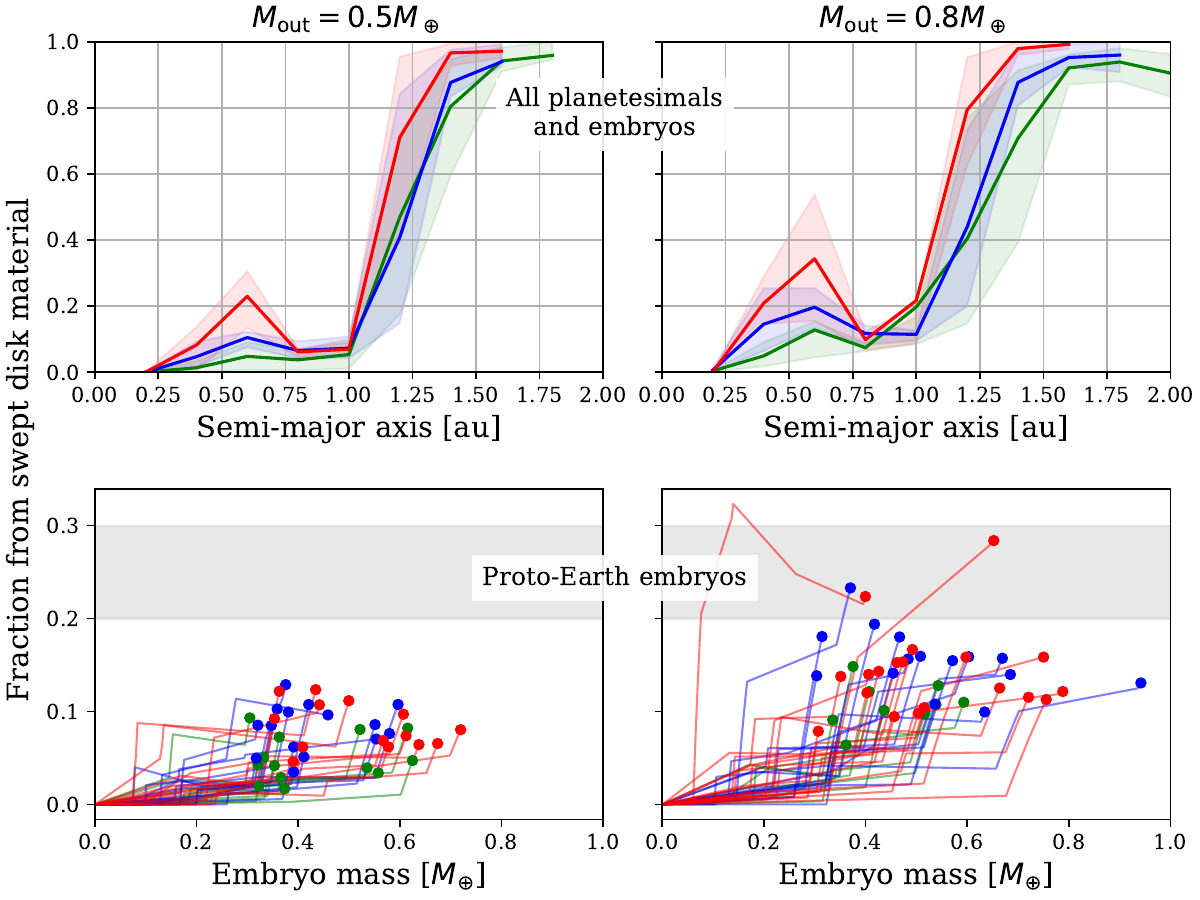}
    \caption{Top: composition of all material (embryos and planetesimals) at the end of the simulation across the terrestrial planet region, described by the fraction that originated in the swept disk. The lines and shaded regions represent averages and $1\sigma$ deviations over five runs per parameter set and two Jupiter-Saturn configurations. Green, blue, and red are $\tau_\textrm{diss}=$ 1, 2, and 3 Myr, respectively. Bottom: composition during growth of embryos that ended with $\qty{0.85}{au}<a<\qty{1.1}{au}$ and $m>\qty{0.3}{M_\oplus}$, which we consider proto-Earths. Most simulated embryos first accrete inner disk and unswept disk material, then switch to material from the swept disk for the final 10--20\% of their growth. The shaded region indicates the estimated fraction of Earth's accretion that consisted of ordinary chondrite-like material \citep{Dale2025a}.}
    \label{fig:composition}
\end{figure*}

The final state of our simulations has a strong compositional gradient as a function of heliocentric distance. Figure~\ref{fig:composition} shows the mass fraction of embryos and planetesimals that originated from the swept disk across the terrestrial planet region. The fraction ranges from $\approx 10\%$ in the Mercury region to $\approx 90\%$ near the current location of Mars. Let's hypothesize that the inner ring and unswept disk is reduced and that the swept disk material is oxidized. Then the small embryos stranded near 0.5 au are highly reduced, similar to the composition of Mercury. The embryos that formed in the outer ring after it concentrated during the secular resonance are oxidized, similar to Mars.

The large, central embryos are more complicated. We will focus on Earth because oxidation constraints are not available for Venus. We define proto-Earth embryos to be those with $0.85<a<1.1$ and $m>\qty{0.3}{M_\oplus}$ at the end of the simulation. In our calculations, these embryos consistently undergo heterogeneous accretion in which they initially accrete only reduced material and then switch to mostly oxidized material, as seen in Figure~\ref{fig:composition}. Such heterogeneous accretion is the scenario envisioned by \cite{Rubie2011} and \cite{Dale2025a} to reproduce both the silicon and iron oxide abundances in Earth's mantle. At the end of our simulations, the swept disk contribution accounts for 5--20\% of proto-Earth embryos' mass, depending on the choice of parameters. The largest values are seen for higher $M_\textrm{out}$ and longer $\tau_\textrm{diss}$. 


A comprehensive and quantitative comparison to geochemical constraints will require simulations that reach the end of terrestrial planet accretion. \cite{Rubie2011} estimated that the late oxidized accretion accounts for 30--40\% of Earth's mass, but \cite{Dale2025a}, who used an improve differentiated model, reduced this to 20--30\% depending on the exact accretion history. While few of the proto-Earths in Figure~\ref{fig:composition} exceed 10\%, they are rapidly accreting swept disk planetesimals at the end of the simulation. Furthermore, several Mars-sized embryos made almost entirely of swept disk material are frequently present between 1 and 2 au (see Figure~\ref{fig:avse}). Although one of them must survive to become Mars, the rest are likely to impact Earth during the giant planet instability and the final one could in fact be Theia \citep{Jacobson2025}.

\subsection{Unaccreted planetesimals}
Some particles reach the end of the simulation as superparticles, i.e., they did not experience any collisions in 15 Myr. Table~\ref{tab:stats} shows the fraction of initial superparticles that remain unaccreted broken down by their original location. Of superparticles originating in the inner ring, almost all accrete onto an embryo and no more than 3 remain in most cases. In contrast, a large fraction of the superparticles originating beyond 1 au are not accreted. Particles swept by the secular resonance are untouched 20\% to 70\% of the time, and unswept superparticles remain abundant even though most of them are scattered below 1 au. These unaccreted superparticles represent populations of planetesimals that remain available to be accreted onto embryos, deliver a late veneer on the final planets, or be transported to other locations and hence have important implications for the chemical structure of the Solar System. Notably, these planetesimals suffer a very severe collision environment \citep{Bottke2006,Shuai2025} so that, rather than original planetesimals, they should be thought of as collisional fragments. We discuss the plausible role of each planetesimal population below.


\section{Discussion}
\begin{figure*}
    \centering
    \includegraphics[width=0.8\linewidth]{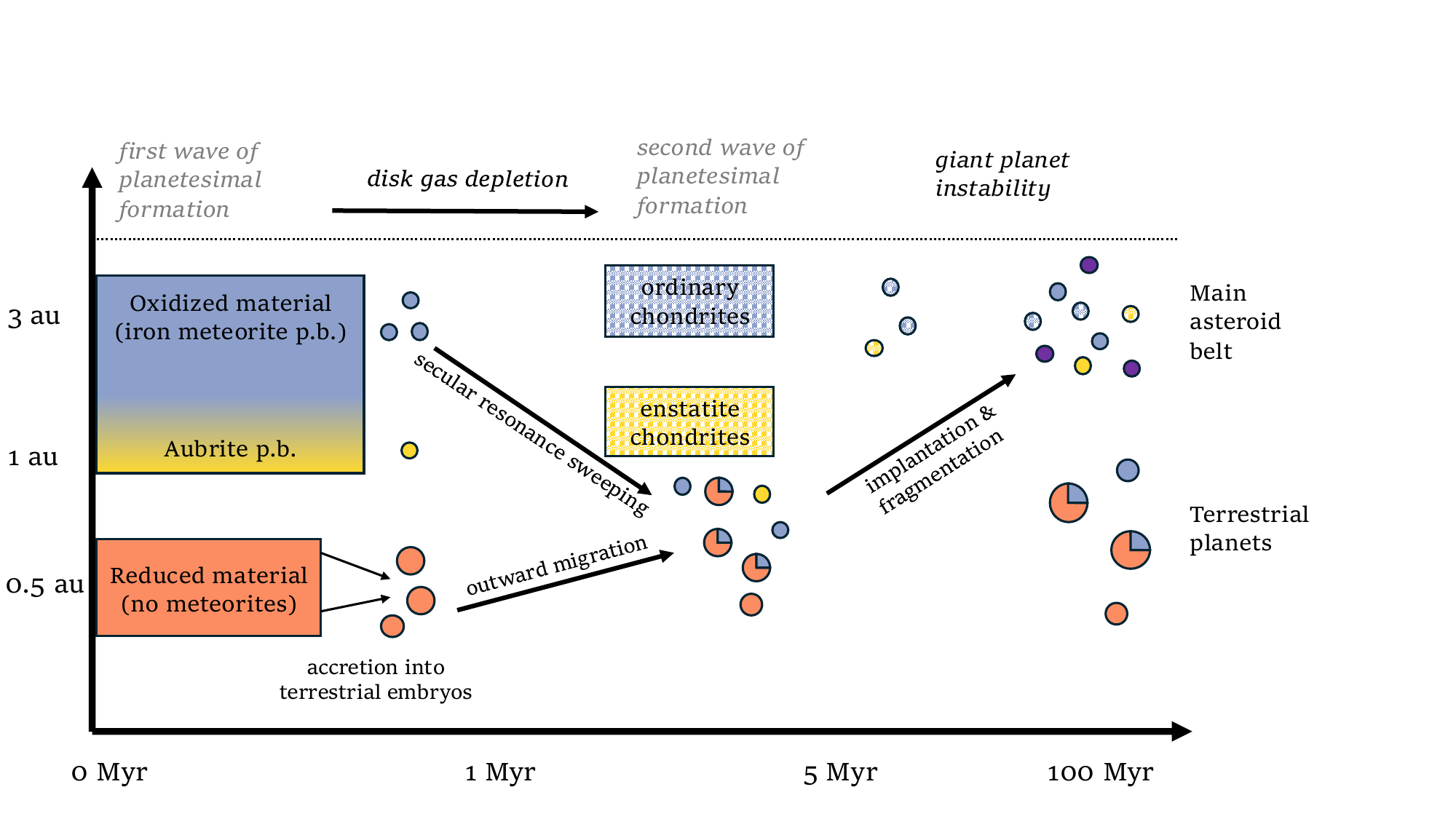}
    \caption{A sketch of the proposed scenario. Oxidized planetesimals decay to near 1 au, accreting onto proto-Earth and proto-Venus as well as forming Mars. Some oxidized planetesimals and aubrite-like planetesimals remain and will be implanted during the giant planet instability into the asteroid belt as the parent bodies of NC iron meteorites and aubrites, where they join the later-formed chondrites (checkered circles) and C-type asteroids (purple circles) delivered from beyond Jupiter.}
    \label{fig:scenario}
\end{figure*}

Figure~\ref{fig:scenario} gives a graphical overview of our proposed scenario of material accretion and transport in the inner Solar System. We envision that an early wave of planet formation makes rocky planetesimals at several locations within the inner disk \citep{Marschall2023} that are differentiated due to high \ce{^{26}Al} abundances. A narrow ring near $\sim 0.5$ au emerges due to dust pileup at the silicate sublimation line and the resulting planetesimals are chemically reduced and refractory enriched, while a more dispersed population appears beyond 1 au whose oxidation state ranges from that of aubrites (reduced) to ordinary chondrites (oxidized). The inner ring accretes quickly into embryos which migrate outwards to the pressure maximum at 1 au while the outer population is relatively static. Meanwhile, Jupiter and Saturn grow to their near-final masses and capture into 2:1 mean-motion resonance. As the gas in the inner disk depletes, the outer planetesimals enter a secular resonance with Jupiter and migrate inwards by aerodynamic drag, reaching the 1--1.5 au region. Many are accreted onto the largest embryos that will become Earth and Venus but another purely oxidized embryo is formed and will become Mars. The gas dissipation also triggers another wave of planetesimal formation, probably near 1 and 2--3 au from remaining dust untouched by the secular resonance; these are the enstatite and ordinary chondrites, respectively. Sometime after disk dissipation, the giant planets undergo an instability and excite orbits near the terrestrial planets. At this time, the largest embryos merge and become the terrestrial planets while a small fraction of objects are implanted into the asteroid belt; these are the aubrite, enstatite, and iron meteorite parent bodies, as detailed below.

The complexity of this hypothesis appears to be justified by the diversity of meteorite compositions and constraints from terrestrial planet embryos that need to be respected. A much simpler scenario does not seem feasible. We justify this scenario and discuss a few points in more detail in this section.

\subsection{Inferred primordial chemical and isotopic composition}
The dynamical behavior of the different planetesimal populations in our simulations combined with the chemical composition of Solar System material give few options for the identity and nature of the initial material. Particles either end up accreted onto embryos that can grow into terrestrial planets or are left as superparticles, representing a large unaccreted planetesimal population. The former category leaves its mark on the compositions of terrestrial planet mantles, for which we will consider the oxidation states of Mercury, Earth, and Mars, and the nucleosynthetic isotopic properties of Earth and Mars. The latter category of remaining planetesimals is important because during the giant planet instability, excitation from Jupiter can deliver material from the terrestrial planet region into the asteroid belt \citep{Raymond2017a}. The probability that a planetesimal present just before the instability will be implanted is small, typically of the order $10^{-4}-10^{-3}$ \citep{Izidoro2024}. This implies that planetesimals that broke up in thousands of fragments are more likely to be sampled in the asteroid belt than planetesimals that survived intact despite the highly collisional environment. Thus, unlike the ordinary chondrites, which formed in situ in the asteroid belt, objects implanted back into the asteroid belt from the terrestrial planet region should only be fragments of original planetesimals \citep{Bottke2006}.

\subsubsection{Inner ring}
\label{sec:inner}
In our simulations, inner ring planetesimals form the vast majority of the innermost embryos near the location of Mercury, and the first 80--90\% of the larger proto-Earth embryos by mass. These planetesimals therefore must have been highly reduced and refractory-enriched to match the inferred compositions of both Mercury \citep{Nittler2011} and the material that comprised the first stage of accretion onto Earth \citep{Rubie2011,Dale2025,Dale2025a}. As the dominant building blocks of Earth---an end member in many isotopic systems---their isotopic composition must have also been the ``lost'' material proposed by \cite{Burkhardt2021}. Indeed, accretion of planetesimals in the inner ring is very efficient, leaving only 0--3 superparticles by the end of the simulation (Table~\ref{tab:stats}), corresponding to $\lesssim 10000$ planetesimals of 100 km diameter from this population. \cite{Nesvorny2025} found from merged simulations that less than one in $10^7$ of the original planetesimals find their way into the asteroid belt by the end of terrestrial planet formation, making this inner ring effectively unsampled in the current asteroid and meteorite collection.

\subsubsection{Swept disk}
\label{sec:swept}
Planetesimals in the swept disk remain mostly undisturbed until $t\approx 2\tau_\textrm{diss}$, when the sweeping secular resonance concentrates them into a narrow ring. The outermost proto-Mars embryos at the end of the simulation are made entirely of swept disk material (Figure~\ref{fig:avse}) while the large proto-Earth and proto-Venus embryos near 1 au receive a small contribution. The Martian mantle is isotopically close to ordinary chondrites and known to be relatively oxidized, with a Mg/Si value similar to ordinary chondrites \citep{Yoshizaki2020}. The final 20--30\% of Earth's accretion, captured in our simulations up to 15 Myr and continuing after from remaining planetesimals and giant impacts, must also have been primarily of oxidized material similar in composition to ordinary chondrites \citep{Rubie2011,Dale2025,Dale2025a}. Together, these constraints imply that the swept disk material must have been isotopically and compositionally similar to ordinary chondrites, that is, NC and relatively oxidized.

Remnants of swept disk material should have been thoroughly implanted into the asteroid belt. About three orders of magnitude more planetesimals from the swept disk are unaccreted at 15 Myr than those from the inner ring. This is because efficient accretion only occurs during the phase when the swept material constitutes a narrow and dense ring, but the ring soon spreads viscously after exiting the secular resonance (see Figure~\ref{fig:tracks}). Furthermore, most of these planetesimals are likely to have been broken \citep{Bottke2006,Shuai2025}. They formed early, before the secular resonance sweeping, and therefore should have differentiated under the effect of \ce{^{26}Al} decay. Given the high collision velocities involved during the secular resonance and once in the terrestrial planet region \citep[$> \qty{10}{km.s^{-1}}$;][]{Bottke2006,Shuai2025}, it is likely that collisional disruptions liberated debris of their cores.

Thus, following \cite{Shuai2025}, we propose that the debris of the cores of these swept disk planetesimals that have been re-captured into the asteroid belt are the parent bodies of the NC iron meteorites that we receive today on Earth. NC irons are indeed isotopically similar to ordinary chondrites and their chromium, nickel, and cobalt abundances imply that the parent bodies were relatively oxidized \citep{Bonnand2018,Grewal2024}. Furthermore, the essentially random implantation of a large number of fragments explains why the iron meteorite sample appears to come from a large number of distinct parent bodies \citep{Spitzer2025}.

\subsubsection{Unswept disk}
The planetesimals that originated between 1 and 1.5 au do not experience rapid secular resonance-assisted radial migration but still gradually drift inwards by aerodynamic drag. Their evolution is determined by the largest embryos, which accrete them or scatter them inwards upon reaching 1 au. A moderate fraction survive to 15 Myr unaccreted and could be implanted into the asteroid belt. If we assume a continuous oxidation gradient in the inner Solar System, the unswept disk material should have an oxidation state between the highly reduced inner ring and the oxidized swept disk. We propose therefore that aubrites, which are reduced and isotopically nearly identical to enstatite chondrites but fully differentiated \citep{Keil1989}, sample the unswept disk. Aubrites and enstatites could not come from the inner ring because they do not possess the end-member isotopic and chemical properties for the inner ring population (see Section~\ref{sec:inner}).

\subsection{Chondrites}
The NC chondrites accreted later, typically $\sim \qty{2}{Myr}$ after the formation of Ca-Al-rich inclusions (CAIs) \citep{Kleine2020}, but they retain a gradient in refractory enrichment and oxidation state that we require in our differentiated planetesimals \citep{Krot2014}. This implies that the same dust reservoir that produced the unswept aubrites later formed the enstatite chondrites, and that the reservoir that produced swept NC irons later formed ordinary chondrites. 
Achieving this requires holding dust reservoirs for several Myr without significant mixing. The sweeping secular resonance resolves this conundrum because it does not trigger radial migration of well-coupled dust, and so the first generation of differentiated planetesimals would be removed from the dust ring but leave behind the dust. The dust itself could be trapped in the several weak pressure bumps induced by a growing Jupiter \citep{Lega2025,Srivastava2025}. Thus it is conceivable that the chondrites formed in situ after the belt was cleared by the sweeping secular resonance, a hybrid scenario compared to a completely empty belt filled during the instability \citep{Raymond2017a}. In order to form after secular resonance sweeping, the dissipation timescale of the disk should have been $\sim \qty{1}{Myr}$ (see Figure~\ref{fig:secreslocs}). We exclude that ordinary chondrites were swept by the secular resonances along with the earlier generation of planetesimals because the collisional histories of chondritic asteroids and iron meteorites are radically different.

\subsection{CC contribution to terrestrial planet mantles}
A result of the sweeping secular resonance model presented in this paper is that the final 20--30\% of Earth's accretion should predominantly be material that was present at 1--4 au after the disk gas surface density decayed to $\sim 100 - \qty{500}{g.cm^{-2}}$. As envisioned in this work, this process is needed to explain the iron and silicon oxide abundances of Earth's mantle \citep{Rubie2011} and succeeds if the swept material is similar in composition to ordinary chondrites \citep{Dale2025a}. However, other material could exist in this region and accrete onto Earth. Potentially problematic is the presence of CC material scattered in by the growing Jupiter and Saturn \citep{Raymond2017}, which need to be almost at their final mass for the sweeping secular resonance to proceed. Isotopic evidence of refractory elements demonstrates that Earth and Mars contain no more than 15\% CC material \citep{Burkhardt2021}. 

However, this is not so severe of a problem. First, all known CC chondrites formed after ordinary chondrites \citep{Kleine2020}, so our above constraint that the clearing of the belt during the sweeping secular resonance occurred before the formation of the ordinary chondrites also means that the CC chondrites must have been scattered into the belt after the resonance. Second, the CC iron meteorites could have been implanted before the resonance---indeed, collisional evolution during and after the sweeping could have broken them---as long as they comprised $\lesssim 15\%$ of the proto-asteroid belt, thus contributing $\lesssim 15\%$ and $\lesssim 5\%$ to Mars and Earth, respectively.

\subsection{Timing of the sweeping secular resonance}
The above constraints from chondrites and iron meteorites imply that the secular resonance passed through the proto-asteroid belt region between 1 and 2 Myr after the formation of CAIs, corresponding to $\tau_\textrm{diss}=\qty{1}{Myr}$ in our setup. While such short dissipation timescales typically result in less favorable orbital architectures and contribution of oxidized material to proto-Earth compared to larger values of $\tau_\textrm{diss}$, it is important to note that the timing of the secular resonance and the location of the final ring depends on the details of the gas disk profile, which we adopted without modification from \cite{Nesvorny2025} because of its success in reproducing the architecture of the first three planets via Type I migration. Even a slightly modified surface density profile, such as a broken power law (dash-dotted line in Figure~\ref{fig:secreslocs}), can deliver material earlier and closer to 1 au while still providing convergent migration.

The timing of the sweep is also linked to the formation of Mars because its accretion is triggered directly by the concentration into a ring. In our simulations with $\tau_\textrm{diss}=\qty{1}{Myr}$, Mars typically reaches half of its final mass at $\sim\qty{3}{Myr}$ and its full mass around $\sim \qty{6}{Myr}$. This is slightly slower than the timescales inferred from Hf-W chronology if the Martian mantle is homogeneous \citep{Dauphas2011} but is consistent if there are slight heterogeneities due to large projectiles \citep{Marchi2020}. Rapid formation of Mars while nebular gas was still present also ensures that it could accrete a large primordial atmosphere and thus retain a solar signature in Xe \citep{Joiret2025}.

\section{Conclusion}
In this work, we have performed a suite of N-body simulations modeling the accretion of the terrestrial planets accounting for collisions, viscous stirring, Type I migration, aerodynamic drag, the gas disk potential, and perturbations from the giant planets with the particular aim of identifying the origins of the chemical architecture of the inner Solar System. Expanding on the work of \cite{Woo2023,Woo2024} and \cite{Nesvorny2025}, who demonstrated that convergent Type I migration towards 1 au and a two-ring model are the best solution to the origin of the orbital architecture of the terrestrial planets, our simulations use a physically realistic 2:1 mean-motion resonant configuration of Jupiter and Saturn and naturally account for a sweeping secular resonance that causes long-range migration of planetesimals by aerodynamic drag. 

The key result of this work is that a primordial composition gradient, rapidly erased by mixing in the standard ring model, can be maintained if planetesimals originating outside of 1 au were delivered later during the sweeping secular resonance. The dominant building blocks of the first three planets, necessarily highly reduced and refractory-enriched, originated closer to the Sun and were completely consumed during embryo growth \citep{Nesvorny2025}, in agreement with the ``lost'' reservoir hypothesis \citep{Burkhardt2021}. Then, a second and outer ring formed of planetesimals swept by the secular resonance delivered oxidized proto-asteroid belt material to the terrestrial planet region. These differentiated planetesimals had compositions similar to ordinary chondrites and contributed to the final 20--30\% of Earth's accretion, most of Mars's mass, and were implanted in the asteroid belt as the NC iron meteorite parent bodies. Full verification of this model will require more complete simulations including the giant planet instability, explicit calculation of core-mantle equilibration in the largest embryos, and tracking isotopic evolution.

\begin{acknowledgments}
We thank the referee for useful comments. M.G.~and A.M.~acknowledge support from the ERC project N. 101019380 ``HolyEarth.'' The work of D.N.~was supported by the NASA SSERVI program. This project was provided with computer and storage resources by GENCI at IDRIS thanks to the grants 2024-AD010413416R1 and 2025-AD010413416R2 on the V100 partition of the Jean Zay supercomputer.
\end{acknowledgments}

\bibliography{TerrestrialFormation}
\bibliographystyle{aasjournalv7}

\end{document}